# The Key Steps and Distinct Performance Trends of Pyrrolic *vs*. Pyridinic M-N-C Catalysts in Electrocatalytic Nitrate Reduction

Qiuling Jiang [a, c, d, e], Mingyao Gu [b, e], Tianyi Wang [a], Fangzhou Liu [b], Xin Yang [b], Di Zhang [a], Zhijian Wu [c, d], Ying Wang [c, d, *], Li Wei [b, *], Hao Li [a, *]

- a. Advanced Institute for Materials Research (WPI-AIMR), Tohoku University, Sendai, Japan, 980-0807
- b. School of Chemical and Biomolecular Engineering, The University of Sydney, Darlington, New South Wales, Australia, 2006
- c. State Key Laboratory of Rare Earth Resource Utilization, Changchun Institute of Applied Chemistry, Chinese Academy of Sciences, Changchun, China, 130022
- d. School of Applied Chemistry and Engineering, University of Science and Technology of China, Hefei, China, 230026
- e. These authors contribute equally

**Email:**

ywang_2012@ciac.ac.cn (Y.W.)

l.wei@sydney.edu.au (L.W.)

li.hao.b8@tohoku.ac.jp (H.L.)

**ABSTRACT**

Electrochemical nitrate reduction reaction ($NO_3RR$) offers a sustainable route for ambient ammonia synthesis. While metal-nitrogen-carbon (M-N-C) single-atom catalysts have emerged as promising candidates for $NO_3RR$, the structure-activity relations underlying their catalytic behavior remain to be elucidated. Through systematic analysis of reported experimental data and pH-field coupled microkinetic modelling on a reversible hydrogen electrode (RHE) scale, we reveal that the coordination-dependent activity originates from distinct scaling relations governed by metal-intermediate interactions. M-N-Pyrrolic catalysts demonstrate higher turnover frequencies for ammonia production, whereas M-N-Pyridinic catalysts exhibit broader activity ranges across the activity volcano plot. Meanwhile, the adsorption and protonation of nitrate, which is a step often dismissed and/or assumed to be simultaneous in many previous reports, is identified to be the rate-determining step (RDS) in $NO_3RR$. Remarkably, our subsequent experimental validation confirms the theoretical predictions under both neutral and alkaline conditions. This study offers a comprehensive mechanistic framework for interpreting the electrocatalytic activity of M-N-C catalysts in $NO_3RR$, showing that a classical thermodynamic "limiting-potential model" is not sufficiently accurate to capture the RDS and the catalytic performance trends of different materials (even on M-N-Pyrrolic and M-N-Pyridinic catalysts). These findings provide brand new insights into the reaction mechanism of $NO_3RR$ and establish fundamental design principles for electrocatalytic ammonia synthesis.

## Introduction

The Haber-Bosch process, a cornerstone of industrial ammonia production, demands considerable energy input under high temperature and pressure conditions. The significant energy requirements have prompted research into sustainable alternatives, especially ambient-condition electrochemical processes. Recent electrochemical ammonia synthesis efforts have primarily focused on the nitrogen reduction reaction (NRR), but significant kinetic limitations remain a major challenge. To overcome these constraints, the electrochemical reduction of nitrate ($NO_3RR$) has emerged as a promising alternative route, offering more favorable thermodynamics and kinetics compared to direct NRR.[1]

Within the broader landscape of $NO_3RR$ electrocatalysts, metal-nitrogen-carbon (M-N-C) single-atom catalysts are regarded as potential candidates due to their well-defined atomic structure and superior catalytic activity. The atomic dispersion of metal centers with nitrogen-doped carbon frameworks provides an ideal platform for rational catalyst design through precise control of metal active sites, coordination environments, and functional groups, thereby modulating the electronic structure and catalytic performance.[2] For instance, under alkaline conditions at -0.5 V *vs.* reversible hydrogen electrode (RHE), a pyridinic-N-rich copper single-atom catalyst (PR-CuNC) achieved an $NH_3$ yield of 3.74 mg h$^{-1}$ cm$^{-2}$ with a Faradaic efficiency of 94.61%, significantly higher than those of a low pyridinic-N-coordinated Cu catalyst (CuNC, 1.10 mg h$^{-1}$ cm$^{-2}$, 65.24%).[3]

Currently, the experimental development of M-N-C catalysts mainly relies on empirical trial-and-error approaches. While this experimental strategy has led to substantial advances, it lacks systematic guidance for catalyst design, resulting in inefficient screening processes. To address the limitations of empirical approaches, density functional theory (DFT) calculations are a powerful tool to guide rational catalyst design, offer atomic-level insights into reaction mechanisms, and enable rapid screening of potential catalysts.[4] However, current theoretical models mainly employ thermodynamic descriptors, such as overpotential (η) and limiting potential ($U_L$), to predict catalytic activity. Although these approaches have successfully described several electrochemical processes, including oxygen reduction reaction (ORR),[5] oxygen evolution reaction (OER),[6] and nitrogen reduction reaction (NRR)[7] on some metal and metal oxide catalysts, they inadequately capture the complex pH-dependent catalytic behavior in the RHE scale.

Recent work[8] by Nørskov and colleagues demonstrated that incorporating electric field effects in theoretical microkinetic models can effectively capture pH-dependent catalytic behavior in the RHE scale, thereby improving the accuracy of activity predictions with experimental conditions. This method allows for a precise description of the interfacial electric field at the applied potential. The electric field, which influences the electric double-layer structure, is also related to the potential of zero charge (PZC). To date, this electric-field-based approach has been successfully applied in theoretical studies of oxygen reduction reaction (ORR)[9] and carbon dioxide reduction reaction (CO$_2$RR),[10] benchmarking well with their corresponding experimental observations. However, exploring NO$_3$RR behavior under electric fields remains very limited. Electric fields can significantly alter the adsorption behavior of reaction intermediates, consequently influencing the overall reaction mechanisms. Similar to the weak-bonding characteristics of *O$_2$ and *O intermediates in ORR, NO$_3$RR also involves weak-bonding species, particularly *NO$_3$H and *NO$_2$H, which are often overlooked in conventional theoretical studies[11, 12, 13] due to difficulties in finding stable adsorption configurations without considering electric field effects.

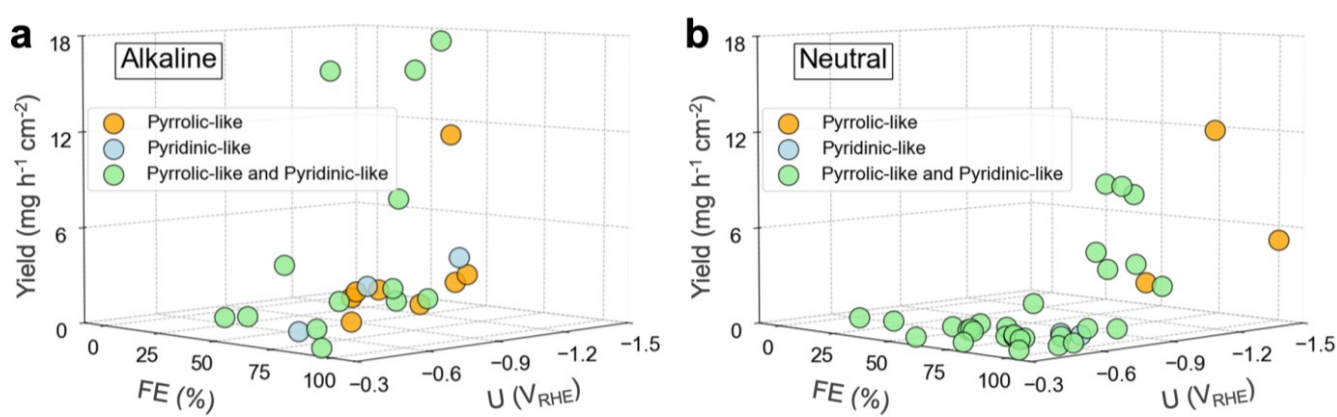

**Figure 1. Analysis of Reported Experimental Performance of Electrocatalytic Nitrate Reduction (NO$_3$RR) to Ammonia on >60 M-N-C Catalysts. (a)** Alkaline conditions. **(b)** Neutral conditions. Detailed information on the M-N-C catalyst performance for NO$_3$RR across the full pH range is provided in **Table S1**. All the source data are available in our Digital Catalysis Platform (*DigCat*) database: https://www.digcat.org/.

Motivated by these challenges, we conduct a comprehensive theoretical investigation of NO$_3$RR on M-N-C catalysts, integrating electric field effects to describe pH-dependent reaction mechanisms under realistic RHE conditions. To begin with, we performed large-scale data mining from the existing experimental literature on NO$_3$RR, and extracted the experimental data of >60 M-N-C catalysts, as summarized in **Figure 1** and **Table S1** (data were also uploaded to the public *DigCat* database:[14] https://www.digcat.org/). Interestingly, the data mining results reveal a predominance of NO$_3$RR studies under alkaline (**Figure 1a**) and neutral conditions (**Figure 1b**), with limited investigations in acidic conditions,[15, 16] mainly due to the competitive hydrogen evolution reaction (HER).[17] Moreover, despite the limited precise characterization of metal coordination environments in reported M-N-C catalysts, which is largely attributed to the high cost of advanced characterization techniques,[18, 19] the available experimental data highlights the superior catalytic activity and Faradaic efficiency of pyrrolic-coordinated M-N-C catalysts under both neutral and alkaline conditions.[20, 21, 22, 23] Based on these observations, we integrated electric field effects into pH-dependent microkinetic modelling and derive activity volcano models using the turnover frequency of ammonia (TOF$_{NH_3}$) as the activity indicator, in contrast to previous theoretical studies[24, 25] based on $U_L$. This precise model reveals distinct pH-dependent activity trends between M-N-Pyrrolic and M-N-Pyridinic catalysts, which is due to the different scaling relations identified on these two different types of M-N-C structures. Furthermore, the theoretical predictions benchmark well with our subsequent experimental validation on structurally well-defined metal phthalocyanine (MPc/CNT) catalysts. In this work, the established pH-dependent structure-activity volcano model (at the RHE scale) not only explains previously observed experimental trends but also provides a foundation for further rational catalyst design in electrocatalytic ammonia synthesis.

**Results and Discussion**

Understanding the adsorption behavior of reaction intermediates and mechanisms of reaction pathways is essential for the rational design of efficient electrocatalysts. To investigate the complex eight-electron proton-coupled electron transfer processes in NO$_3$RR, we first summarized the main reduction pathways on M-N-C catalysts based on available experimental and theoretical evidence.[15, 24] The reaction network is illustrated in **Figure 2**, with detailed elementary steps and three possible reaction routes (**Path 1-3**) provided in the **Supporting Information Section 2.6**.

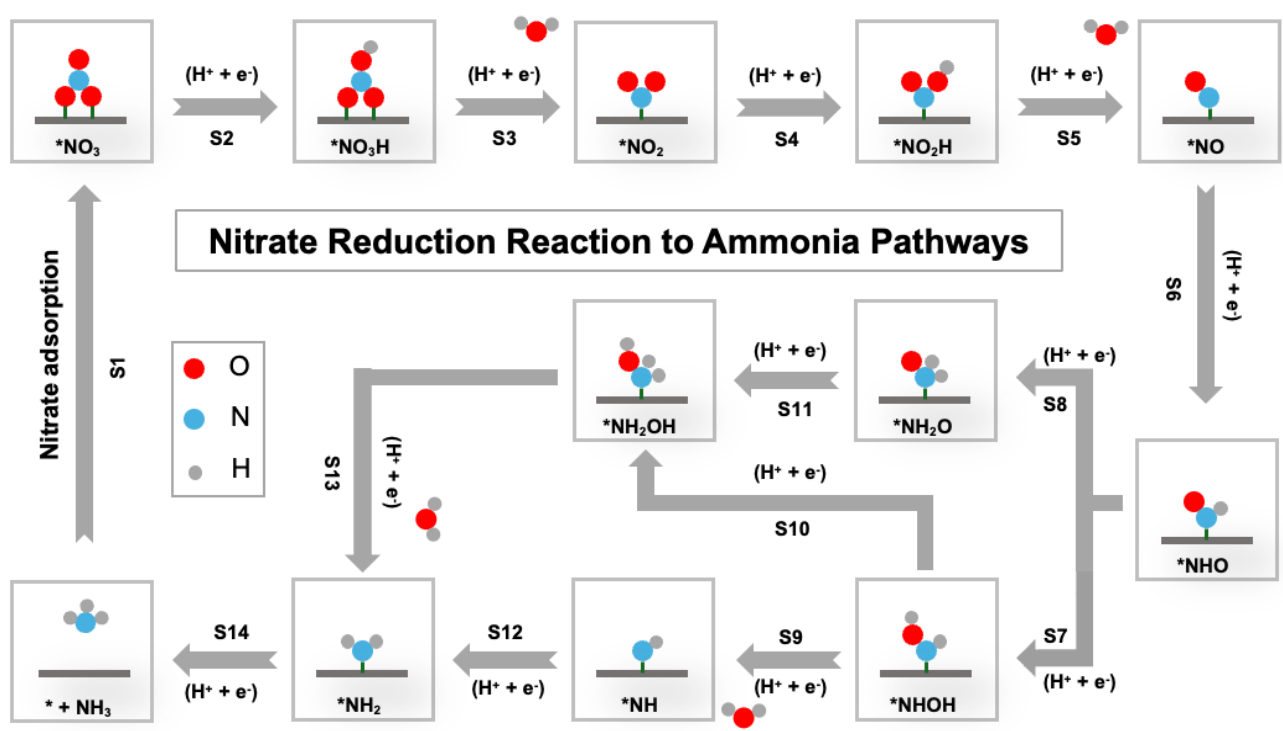

**Figure 2.** Reaction Network of Nitrate Reduction to Ammonia on M-N-C Catalysts.

The reaction network consists of key NO$_3$RR intermediates, and their adsorption behavior on catalysts determines the reaction mechanism. Before elucidating the catalytic behavior of M-N-C catalysts in NO$_3$RR, two primary metal coordination environments were identified (**Figure 3a**). The pyrrolic-like coordination features metal atoms bonded to nitrogen within five-membered rings, while the pyridinic-like coordination involves metal-nitrogen bonds in six-membered rings. According to these coordination environments, four classic M-N-C catalyst structures were investigated: M-N$_4$-Pyrrolic, M-N$_3$-Pyrrolic, M-N$_4$-Pyridinic, and M-N$_3$-Pyridinic. On these M-N-C catalysts, linear scaling relations of intermediate adsorption energies were analyzed. The adsorption free energy of *NH$_2$ ($\Delta G_{ads}$(*NH$_2$)) was identified as the optimal descriptor, with separate fittings for each coordination environment exhibiting higher R² values than the overall fitting for all M-N-C catalysts presented in **Figure S1**. Moreover, **Figures 3b-3j** display the scaling relations between the adsorption free energies of various intermediates ($\Delta G_{ads}$(*NO$_3$), $\Delta G_{ads}$(*NO$_2$), $\Delta G_{ads}$(*NO$_2$H), $\Delta G_{ads}$(*NO), $\Delta G_{ads}$(*NHO), $\Delta G_{ads}$(*NHOH), $\Delta G_{ads}$(*NH$_2$O), $\Delta G_{ads}$(*NH$_2$OH), $\Delta G_{ads}$(*NH)) against $\Delta G_{ads}$(*NH$_2$). During the linear scaling analysis, two important considerations should be noted in this study. Firstly, the scaling relation between $\Delta G_{ads}$(*NO$_3$H) and $\Delta G_{ads}$(*NH$_2$) was not presented due to the physical adsorption of *NO$_3$H on M-N-C catalysts, resulting in negligible adsorption free energy variations among different M-N-C catalysts. The $\Delta G_{ads}$(*NO$_3$H) ranges from -0.08 to -0.10 eV for M-N-Pyrrolic catalysts and from -0.02 to -0.13 eV for M-N-Pyridinic catalysts. Secondly, some anomalous data were excluded from the fitting process, as shown in **Figures 3c**, **3e**, and **3j**, due to the different favorable adsorption sites. Specifically, on weak-binding Cu-N$_3$-Pyrrolic catalysts, *NO$_2$, *NO, and *NH preferentially adsorb on the bridge site between the metal and its coordinating carbon (M-C) rather than at the metal top site. Similar experimental and theoretical obversions were reported in ORR intermediates on weak-binding M-N-C catalysts.[9]

Analysis of the linear regression results reveals significant trends in the adsorption strengths of NO$_3$RR intermediates. Notably, the oxygen-containing intermediates in the initial stages of NO$_3$RR, specifically *NO$_3$ (**Figure 3b**), *NO$_2$ (**Figure 3c**), and *NO$_2$H (**Figure 3d**), show weaker adsorption strengths on M-N-Pyrrolic catalysts compared to M-N-Pyridinic catalysts. In contrast, other reaction intermediates exhibit an opposite trend of adsorption strength. This observation differs from the previously reported higher activity of metal sites in pyrrolic-coordinated structures relative to those in pyridinic-coordinated structures.[26] The observed differences in adsorption strengths are crucial in explaining the varying volcano models for these two coordination environments in the following section. Furthermore, an analysis of covariance (ANCOVA) was performed on the scaling relations for all adsorbates in NO$_3$RR (**Table S4**). The consistently high F-values (>100) in all models indicate substantial differences in adsorption free energies between M-N-Pyrrolic and M-N-Pyridinic catalysts. Among these, the linear relation of *NO$_2$H exhibits the highest F-value (1101.45), emphasizing the most obvious difference between pyrrolic-coordinated and pyridinic-coordinated structures for the *NO$_2$H adsorption. The distinct patterns of adsorption strength significantly influence the rate-determining step (RDS) in these coordination environments, which will be discussed later in the kinetic analysis section.

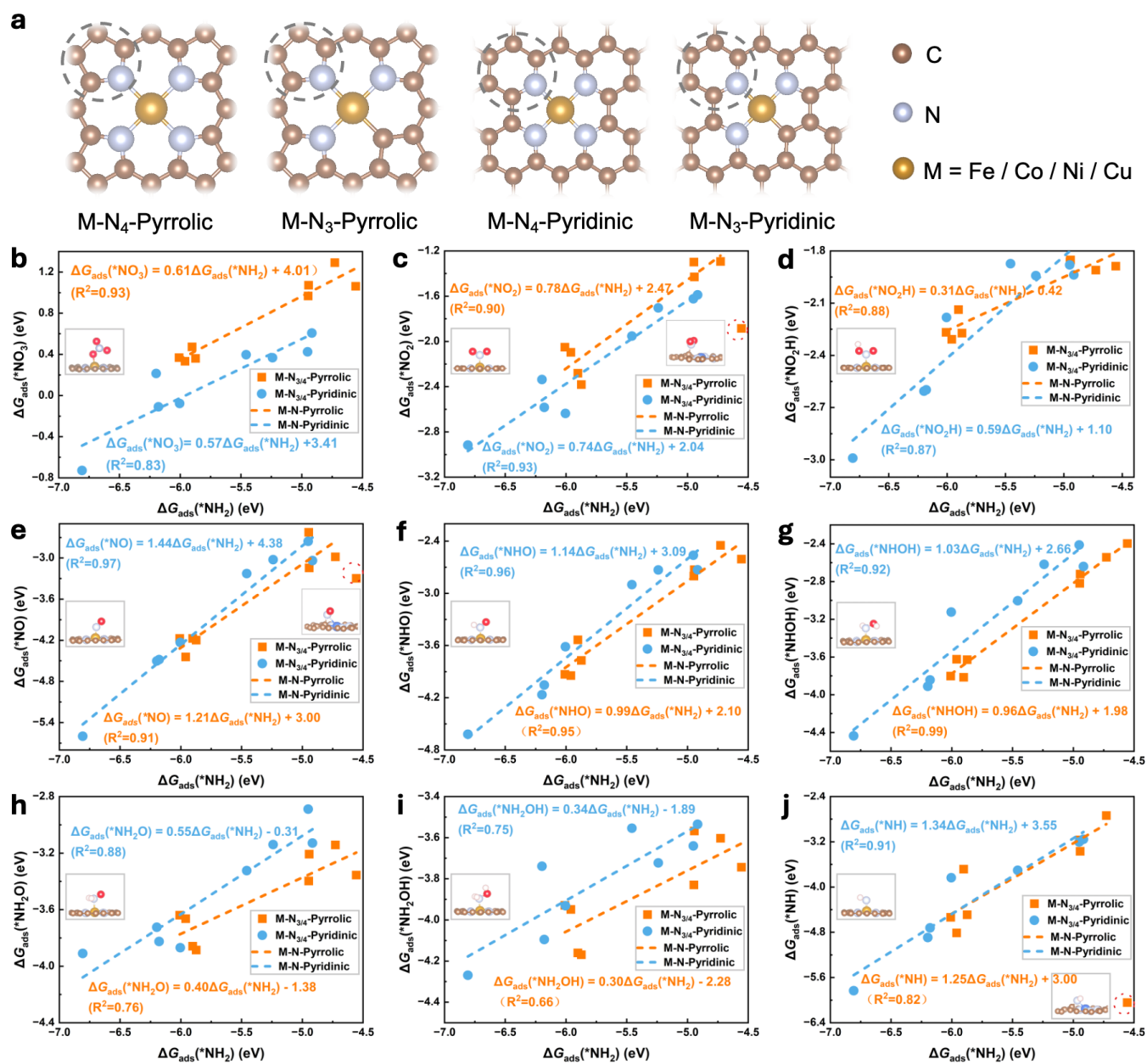

**Figure 3. Structure Configurations and Scaling Relations of the Adsorption Free Energies of NO$_3$RR Intermediates on M-N-C Catalysts.** (a) Schematic structures of M-N$_4$-Pyrrolic, M-N$_3$-Pyrrolic, M-N$_4$-Pyridinic, and M-N$_3$-Pyridinic catalysts. (b-j) Linear scaling relations between $\Delta G_{ads}$(*NH$_2$) and $\Delta G_{ads}$ of various NO$_3$RR intermediates on M-N-Pyrrolic (orange circles) and M-N-Pyridinic (blue squares) catalysts: (b) *NO$_3$, (c) *NO$_2$, (d) *NO$_2$H, (e) *NO, (f) *NHO, (g) *NHOH, (h) *NH$_2$O, (i) *NH$_2$OH, and (j) *NH. Insets: Optimized adsorption configurations of the respective intermediates. Red, brown, light blue, light pink,

and golden spheres represent O, C, N, H, and meta site, respectively.

To enhance the microkinetic model's accuracy under experimental pH conditions and applied potentials under an RHE scale, the potential of zero charge (PZC) with explicit solvent and electric field effects were also incorporated in this study. PZC is fundamental to electrocatalysis simulations, directly correlating with the electric double-layer structure and modulating the interfacial electric field at various applied potentials.[27] Previous studies have indicated that implicit models may have inherent limitations in calculating the PZC when complex water-surface interactions dominate the interfacial region.[28] In this study, PZCs were calculated under explicit solvent conditions to more accurately model the electrode-electrolyte interface. For M-N-Pyrrolic catalysts (M = Fe, Co, Ni, and Cu), $U_{PZC}$ values range from +0.34 to +0.43 $V_{SHE}$ (**Figure 4a**). In contrast, M-N-Pyridinic catalysts exhibit more negative $U_{PZC}$ values between -0.34 and -0.62 $V_{SHE}$ (**Figure 4b**). To maintain consistency in microkinetic modelling, the average $U_{PZC}$ values were applied for catalysts with different coordination environments.

Following PZC calculations, the electric field effect was considered for pH-dependent electrocatalysis simulations. This effect can significantly influence the stability of key intermediates, particularly on weak-binding species.[8] And this approach reveals the interplay between RHE and SHE potential dependencies, facilitating the modelling of pH-dependent activity trends observed experimentally.[27] The integration of electric field effects in microkinetic models enhances the accuracy of catalytic performance predictions across diverse experimental conditions. The field perturbs intermediates and transition states with significant dipole moments (μ) and polarizability (α). **Figures 4d** and **4e** illustrate the fitted response curves for $NO_3RR$ intermediates on Fe-$N_4$-Pyrrolic and Fe-$N_4$-Pyridinic catalysts across an electric field range of -1.0 to +1.0 V/Å, along with the fitted polarizability (α) and dipole moment (μ) values. Electric field response curves for $NO_3RR$ intermediates on other M-N-C catalysts are shown in **Figure S2**. Distinct differences in the electric field response of the same reaction intermediates are observed between M-N-Pyrrolic and M-N-Pyridinic catalysts. This is also particularly evident for the initial oxygen-containing species in the $NO_3RR$, such as *$NO_3$, *$NO_3H$, *$NO_2$, and *$NO_2H$. For instance, *$NO_3H$ under an electric field exhibits a dipole moment of +0.26 eÅ on the Fe-$N_4$-Pyrrolic system and -0.17 eÅ on the Fe-$N_4$-Pyridinic system. These opposing dipole moments result in *$NO_3H$ stabilization under negative fields on Fe-$N_4$-Pyrrolic and positive fields on Fe-$N_4$-Pyridinic, as evidenced by the structural transformations shown in **Figure 4c**. Under varying electric field strengths from +0.6 to -0.6 V/Å, the *$NO_3H$ adsorbate exhibits distinct binding behaviors: the O(*$NO_3H$)-Fe bond distance in the Fe-$N_4$-Pyrrolic system gradually decreases from 3.07 to 1.93 Å, indicating enhanced interaction with the metal site to form a stable adsorption configuration. In contrast, the O(*$NO_3H$)-Fe bond in the Fe-$N_4$-Pyridinic system extends from 1.88 to 1.93 Å, resulting in a destabilized adsorption configuration.

It is noteworthy that *NO$_3$H and *NO$_2$H are often overlooked in theoretical calculations for NO$_3$RR.[11, 13, 29] This omission can be partly attributed to the difficulty in identifying stable structures for *NO$_3$H without considering electric field effects. Some published studies even suggested that *NO$_3$ does not undergo protonation to form *NO$_3$H in the NO$_3$RR process but instead directly dissociates into *NO$_2$ and *O.[25] However, the reported energy barrier for this process is ~1.5 eV even on a strong-binding catalyst surface,[30] suggesting that a direct dissociation of the N-O bond of *NO$_3$ is rather difficult to occur under room temperature. Therefore, considering electric field effects in this study enables a more comprehensive analysis of these crucial intermediates.

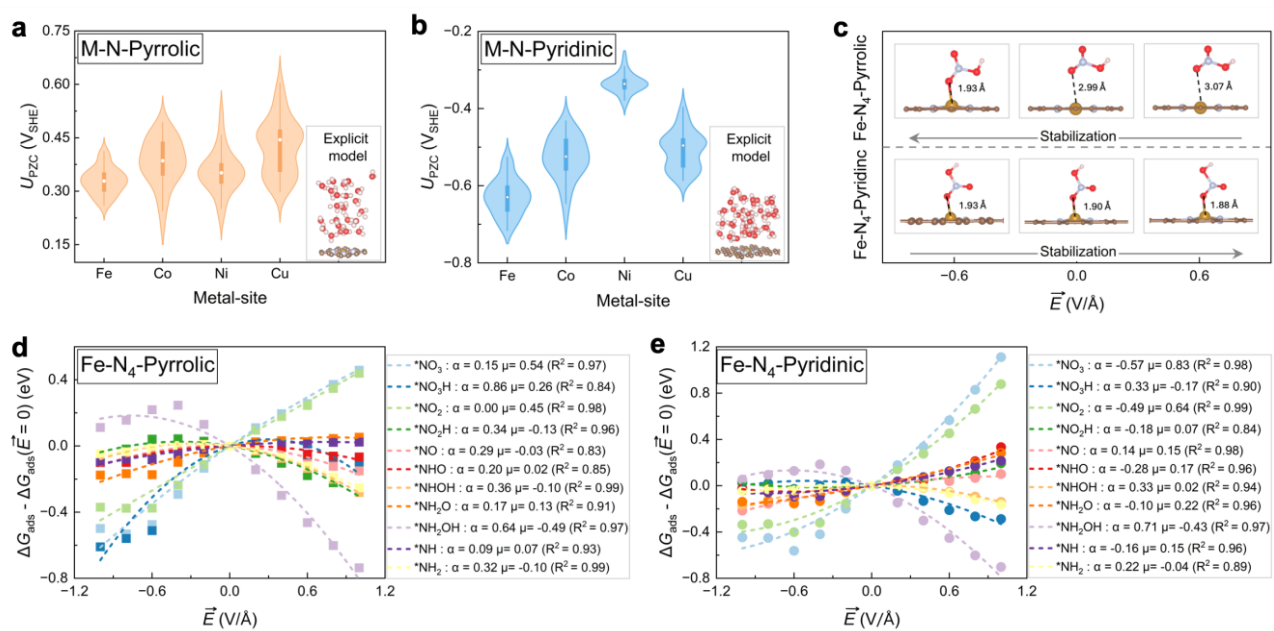

**Figure 4. Potential of Zero Charges (PZCs) and Electric Field Effects.** Calculated $U_{PZC}$ values for **(a)** M-N-Pyrrolic and **(b)** M-N-Pyridinic structures (M = Fe, Co, Ni, and Cu) using the explicit solvent models. The $U_{PZC}$ distributions were derived from >1,000 steps of catalyst-water interfaces from *ab initio* molecular dynamics (AIMD) simulations. Insets: typical configuration of explicit water molecules on Fe-N$_4$-Pyrrolic and Fe-N$_4$-Pyridinic structures. Additional configurations of M-N-Pyrrolic and M-N-Pyridinic structures with explicit water molecules are available in our GitHub and *DigCat* database. Red, brown, light blue, light pink, and golden spheres represent O, C, N, H, and Fe, respectively. **(c)** Optimized *NO$_3$H adsorption configurations on Fe-N$_4$-Pyrrolic (top) and Fe-N$_4$-Pyridinic (bottom) catalysts under applied electric fields. Electric field effects on the adsorption free energies of NO$_3$RR intermediates, with fitted values for polarizability (α, e$^2$ V$^{-1}$) and dipole moment (μ, e Å) for **(d)** Fe-N$_4$-Pyrrolic and **(e)** Fe-N$_4$-Pyridinic catalysts.

Based on a comprehensive analysis of all relevant factors above, more precise microkinetic models for NO$_3$RR on M-N-C catalysts were developed in this study. Full details on the microkinetic modeling are provided in the **Supporting Information**. In contrast to conventional thermodynamic volcano models that utilize limiting potential as an activity indicator (which is often experimentally ill-defined),[24] the microkinetic model in this study employs the experimentally measurable turnover frequency of ammonia ($TOF_{NH_3}$) as the catalytic activity metric. This approach facilitates more meaningful comparisons with experimental data. **Figures 5a** and **5b** display the derived volcano models for $NO_3RR$ at -0.6 $V_{RHE}$ on M-N-Pyrrolic and M-N-Pyridinic catalysts, respectively. For both catalyst types, the volcano model peaks decrease as the pH decreases, indicating higher catalytic activity in alkaline and neutral conditions compared to acidic conditions. This finding aligns with most of the previous relevant experimental results (**Figure 1**), where reports of $NO_3RR$ under acidic conditions are limited, primarily due to two factors: the aforementioned predominance of competing HER and the observed lower $NO_3RR$ activity in acidic conditions. Beyond the general pH-dependent trend, the volcano models also reveal distinct characteristics for both M-N-Pyrrolic and M-N-Pyridinic catalysts. Under the same pH environment, M-N-Pyrrolic catalysts exhibit higher $TOF_{NH_3}$ values, indicating enhanced catalytic activity in specific conditions. Conversely, M-N-Pyridinic catalysts demonstrate a broader range of high activity on the volcano plot, suggesting more consistent performance across diverse conditions. **These contrasting features underscore the critical role of catalyst structure in determining $NO_3RR$ efficiency.**

To gain deeper insights into the microkinetic reaction model, an investigation of the reaction rates of elementary steps in $NO_3RR$ was conducted. **Figures 5c** and **5d** present the RDS analysis for M-N-Pyrrolic and M-N-Pyridinic catalysts, respectively. **Unlike conventional thermodynamic volcano models characterized by a single activity peak, these results exhibit a more complex activity profile with multiple rate-determining regions.** This complexity reveals that three distinct elementary reaction steps govern the catalytic performance rather than the typical two-step process observed in traditional models. Specifically, when the *$NH_2$ adsorption strength is weak, nitrate adsorption serves as the RDS for both catalyst types. As the *$NH_2$ adsorption strength increases, the protonation of *NO to *NHO becomes rate-determining. With a strong *$NH_2$ adsorption strength, the RDS differs for the two catalyst types: for M-N-Pyrrolic catalysts, it is the protonation of *$NO_2$ to *$NO_2H$, while for M-N-Pyridinic catalysts, it is the protonation of *$NO_3$ to *$NO_3H$. These findings indicate that the initial deoxygenation and protonation steps of nitrate predominantly determine the RDS for M-N-C catalysts in $NO_3RR$, consistent with experimental observations.[17, 31]

These variations in RDS can be attributed to the different scaling relations, as shown in **Figures 3b-3j**. For weak *$NH_2$ adsorption strength, the discrepancy in activity ranges between the two catalyst types stems from varying scaling relations for the adsorption free energies of intermediate species *$NO_3$, *NO, and *NHO. This explains the previously mentioned contrasting difference in the width of high-activity regions on the volcano plots. The divergence in RDS at strong *$NH_2$ adsorption can be attributed to the different *$NO_2H$ adsorption behavior. The scaling relations for *$NO_2H$ adsorption free energy ($\Delta G_{ads}$(*$NO_2H$)) relative to *$NH_2$ adsorption free energy ($\Delta G_{ads}$(*$NH_2$)) exhibit significant differences between the two catalyst types. For M-N-Pyrrolic catalysts, the relation is described by: $\Delta G_{ads}$(*$NO_2H$) = 0.31$\Delta G_{ads}$(*$NH_2$) − 0.42 ($R^2$ = 0.88), while for M-N-Pyridinic catalysts, it follows: $\Delta G_{ads}$(*$NO_2H$) = 0.59$\Delta G_{ads}$(*$NH_2$) + 1.10 ($R^2$ = 0.87). These relations indicate that at strong *$NH_2$ adsorption, *$NO_2H$ binds more weakly to M-N-Pyrrolic catalysts compared to M-N-Pyridinic catalysts. Consequently, *$NO_2H$ is more prone to desorption on M-N-Pyrrolic catalysts, potentially forming nitrite, a major by-product of $NO_3RR$. As a result, the formation of *$NO_2H$ emerges as the RDS for M-N-Pyrrolic catalysts in the strong *$NH_2$ adsorption strength regime.

At the fundamental atomic level, these differences can be traced back to the distinct intrinsic properties of M-N-Pyrrolic and M-N-Pyridinic catalysts. For instance, electronic structure analysis of Co-$N_4$-Pyrrolic and Co-$N_4$-Pyridinic catalysts demonstrates the origin of these intrinsic differences, particularly by investigating *$NO_2H$ adsorption behavior. Significant differences in *$NO_2H$ adsorption characteristics between the two catalyst types were observed, as shown in **Figure 5e**. The charge transfer analysis reveals distinct electronic interactions between *$NO_2H$ and the catalysts. Upon *$NO_2H$ adsorption, electron transfer from the catalyst surface to the intermediates facilitates N-O bond activation for subsequent protonation. Specifically, the transferred electron numbers are +0.09 and +0.27 |$e^-$| for Co-$N_4$-Pyrrolic and Co-$N_4$-Pyridinic catalysts, respectively. Crystal orbital Hamilton population (COHP) analysis demonstrates stronger Co-N bonding in Co-$N_4$-Pyridinic catalyst (ICOHP = -2.07) relative to Co-$N_4$-Pyrrolic catalyst (ICOHP = -1.69), indicating enhanced *$NO_2H$ activation on Co-$N_4$-Pyridinic catalyst. Notably, these results deviate from previous studies where pyrrolic-N coordination environments typically exhibit stronger intermediate adsorption and higher catalytic activity relative to Pyridinic-N environments.[26] As shown in **Figure 5f**, *$NO_2H$ demonstrates weaker adsorption ($\Delta G_{ads}$(*$NO_2H$)) on M-N-Pyrrolic relative to M-N-Pyridinic catalysts. Moreover, the correlation between $\Delta G_{ads}$(*$NO_2H$) and metal-N(*$NO_2H$) interaction strength (ICOHP) exhibits distinct linear relations depending on the coordination environment. This behavior originates from the fundamental differences in the electronic structure of the metal centers induced by their coordination environments, which modulates the binding strengths with reaction intermediates and ultimately determines the reaction mechanism in $NO_3RR$.

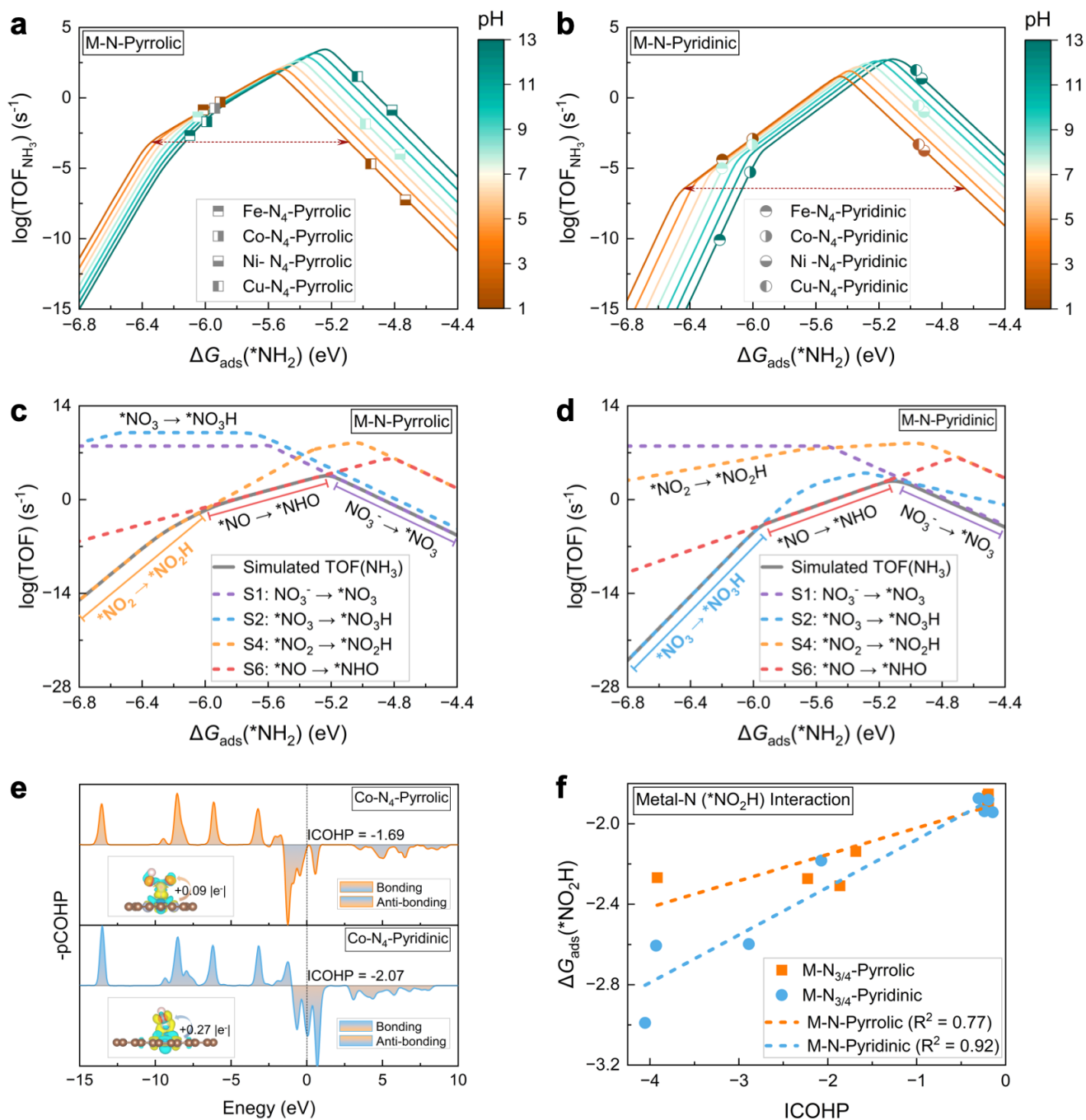

**Figure 5. pH-Dependent Microkinetic Modelling of NO₃RR on M-N-C Catalysts.** pH-dependent activity volcano models for NO₃RR to ammonia at $U$ = -0.6 $V_{RHE}$ on **(a)** M-N-Pyrrolic and **(b)** M-N-Pyridinic catalysts. Rate-determining step (RDS) analyses of NO₃RR in alkaline conditions for **(c)** M-N-Pyrrolic and **(d)** M-N-Pyridinic catalysts. **(e)** pCOHP analysis of the metal site (Co) and N (*NO₂H) interaction on Co-N₄-Pyrrolic and Co-N₄-Pyridinic catalysts. **(f)** The linear relations between integrated COHP (ICOHP) and adsorption free energy of *NO₂H ($\Delta G_{ads}$(*NO₂H)) for M-N-C catalysts.

**This study offers a comprehensive mechanistic framework for interpreting the electrocatalytic activity of M-N-C catalysts in NO$_3$RR, showing that a classical thermodynamic "limiting-potential model" is not sufficiently accurate to capture the RDS and the catalytic performance trends of different materials (even on M-N-Pyrrolic and M-N-Pyridinic catalysts).** To rigorously evaluate its predictive capability and potential for the rational design of high-performance NO$_3$RR electrocatalysts, we carried out subsequent experimental validation using a series of heterogeneous molecular catalysts, which are constructed from depositing metal phthalocyanine molecules on carbon nanotubes (**Figure 6a**, denoted as MPc/CNT, M= Mn, Fe, Co, Ni, and Cu, see experimental details in the **METHODS section**). The main reason for selecting MPc as the subsequent benchmarking analysis is that compared to the pyrolysis synthesis methods (which may generate different degrees of defects and multiple coordination environments), MPc catalysts have well-defined structures after synthesis, and we can consider the exact molecular catalyst model in DFT calculations and the subsequent pH-dependent microkinetic modelling.[32, 33] Inductively coupled plasma atomic emission spectroscopy (ICP-AES) measurement afforded comparable metal loading of ~0.4 wt% (**Table S6**). A representative transmission electron microscope image of the FePc/CNT catalyst is displayed in **Figure 6b** with its corresponding EDX elemental mapping results (**Figure 6c**). Spatially distributed high-contrast spots can be assigned to the Fe atom located in the MPc molecules. No FePc aggregates can be observed. Their corresponding EDX elemental mapping results also exhibit a uniform distribution of C, N, and Fe elements. Other MPc/CNT catalysts exhibit identical morphology under TEM observation (**Figure S3**). High-resolution metal 2$p$ XPS spectra are displayed in **Figure S4**, which resembles the metal center in a +2 valence state. We further collected their XANES spectra and performed EXAFS analysis (**Figure S5**). The single M-N path with a coordination number of 4 found in the first shell for all samples (**Table S7**) confirms the formation of single-atom catalysts with an identical M-N$_4$ structure, allowing us to perform accurate NO$_3$RR performance benchmarking with our theoretical models developed above.

The NO$_3$RR performance was tested in alkaline (0.1 M KOH) and neutral (0.5 M K$_2$SO$_4$) electrolytes, both containing 0.1 M KNO$_3$. The experimental current densities ($j_{Total}$) and turnover frequencies (TOF$_{NH_3}$) of MPc/CNT catalysts under both conditions are presented in **Figures 6d**, **6e**, **6g**, and **6h**, with additional activity data shown in **Figures S6** and **S7**. For comparison, the Faradaic efficiencies of NH$_3$ (FE$_{NH_3}$) at various potentials are also summarized in **Table S8**. These results demonstrate that despite their identical structures, MPc/CNT catalysts exhibit distinct NO$_3$RR activity and NH$_3$ selectivity under different pH conditions. The calculated TOF$_{NH_3}$ values at $U$ = -0.8 V$_{RHE}$ are benchmarked with experimental results in **Figures 6f** and **6i**. The comparison at different potentials (-0.6 and -1.0 V$_{RHE}$) in both alkaline and neutral conditions (**Figure S10**) demonstrates consistent agreement between theory and experiment. Based on this validated model, the detailed free energy diagrams of all NO$_3$RR pathways on these MPc/CNT catalysts (M = Mn, Fe, Co, Ni, and Cu) in alkaline and neutral electrolytes at $U$ = -0.8 V$_{RHE}$ are analyzed in **Figures S11-S16.** At this applied potential, most elementary steps proceed without activation barriers, except for S1: NO$_3^-$ → *NO$_3$, S4: *NO$_2$ → *NO$_2$H, and S6: *NO → *NHO. The main reaction pathways for NH$_3$ synthesis vary with different catalysts and electrolyte conditions (**Figure S11**). In alkaline conditions, NO$_3$RR to NH$_3$ proceeds predominantly through **Path1 (S1→S2→S3→S4→S5→S6→S7→S9→S12→S14)** on MnPc/CNT, while **Path3 (S1→S2→S3→S4→S5→S6→S8→S11→S13→S14)** dominates on other MPc/CNT catalysts (M = Fe, Co, Ni, and Cu). In neutral conditions, NO$_3$RR follows **Path1** on MPc/CNT (M = Mn, Fe, and Co), whereas **Path3** dominates on NiPc/CNT and CuPc/CNT.

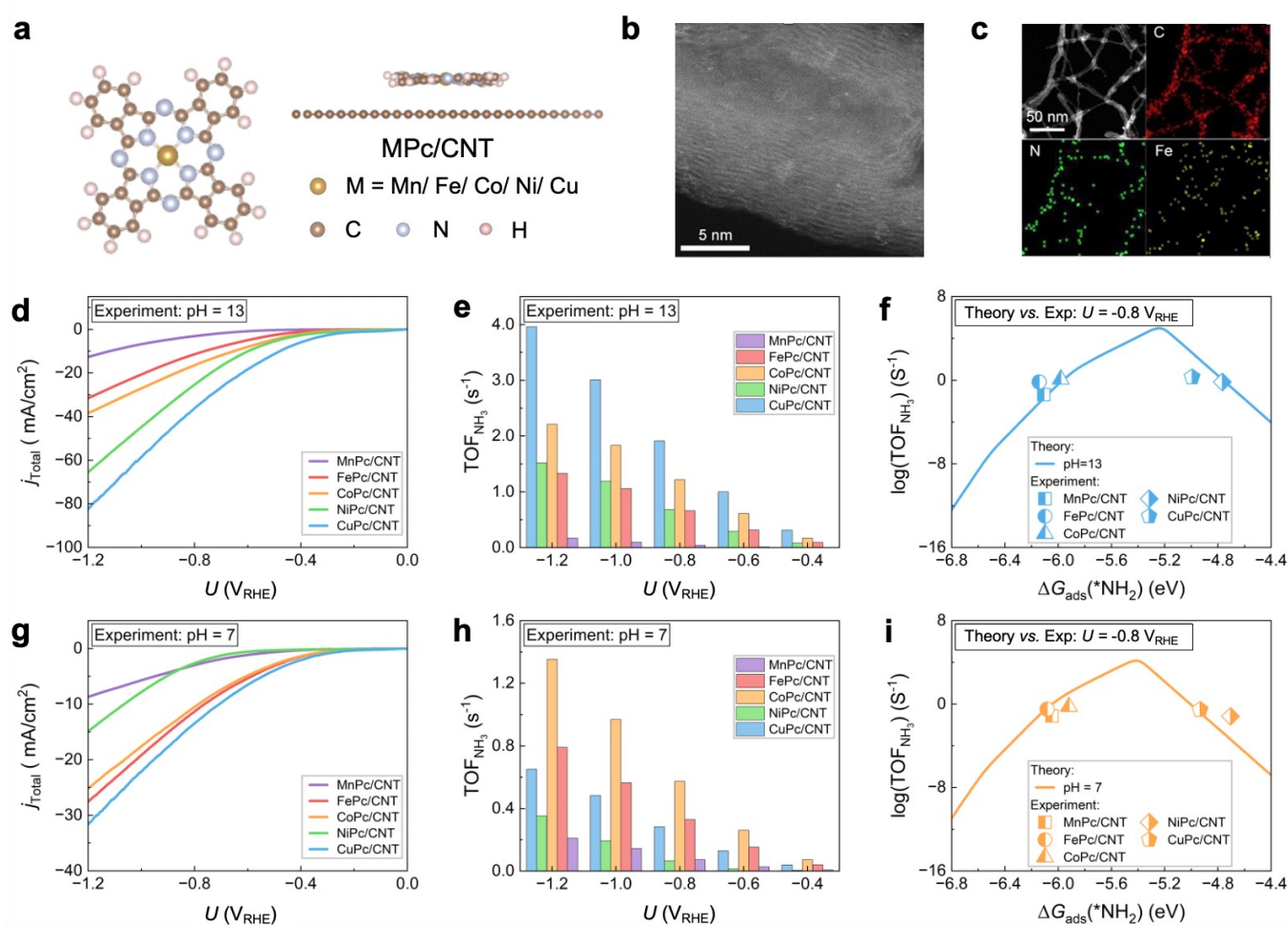

**Figure 6. Benchmarking Analyses Between Theoretical Simulations and Subsequent Experimental Validations. (a)** Atomic structure model of MPc/CNT catalyst. **(b)** HAADF-STEM image of FePc/CNT and **(c)** corresponding EDX elemental mapping. **(d, g)** Experimental current density ($j_{Total}$) during NO$_3$RR under alkaline and neutral conditions, respectively. **(e, h)** Experimental turnover frequency (TOF$_{NH_3}$) under alkaline and neutral conditions, respectively. **(f, i)** Experimental and theoretical TOF$_{NH_3}$ comparison at $U$ = -0.8 V$_{RHE}$ in alkaline and neutral conditions, respectively.

**Conclusion**

This research presents a comprehensive mechanistic framework for understanding the NO$_3$RR on M-N-C catalysts and establishes a precise model that systematically addresses the structure-activity relations underlying their catalytic behavior. **By integrating microkinetic modelling with electric field effect simulations, this study highlights that key elementary steps (such as *NO$_3$→*NO$_3$H and *NO$_2$ → *NO$_2$H) should be explicitly considered rather than being dismissed or treated as simultaneous processes.** Furthermore, the comprehensive microkinetic analysis reveals that M-N-Pyrrolic catalysts exhibit higher TOF$_{NH_3}$ values but narrower activity ranges, whereas M-N-Pyridinic catalysts exhibit broader activity ranges across the activity volcano model. These distinct catalytic behaviors fundamentally stem from their different coordination environments, manifesting through unique scaling relations and rate-determining steps.

These mechanistic insights reveal key strategies for optimizing M-N-C catalysts for NO$_3$RR. Nitrate adsorption and nitric oxide protonation steps are crucial for both coordination environments, particularly when catalysts exhibit weak *NH$_2$ adsorption strengths. For M-N-Pyrrolic catalysts, the focus should be on stabilizing *NO$_2$H to prevent nitrite desorption. In contrast, for M-N-Pyridinic catalysts, attention should be directed toward optimizing the *NO$_3$ activation and protonation steps. The pH-dependent activity trends indicate that alkaline and neutral conditions are more favorable for NO$_3$RR, explaining the historical challenges of realizing efficient NO$_3$RR in acidic conditions.

This theoretical framework establishes clear design principles for promising NO$_3$RR catalysts. Future developments should focus on tailoring coordination environments to achieve optimal binding energetics while considering pH-dependent behaviors. The insights extend beyond NO$_3$RR to benefit other electrochemical reduction reactions where similar coordination effects may play crucial roles. Combined with experimental validation, these findings establish a foundation for the rational design of high-performance M-N-C catalysts.

**Methods**

***Computational methods*** Spin-polarized density functional theory (DFT) calculations were carried out using the Vienna *Ab initio* Simulation Package (VASP).[34, 35, 36] The revised Perdew-Burke-Ernzerhof (RPBE) functional and projector augmented-wave (PAW) method were employed to describe exchange-correlation interactions and core electrons, respectively.[37, 38] A plane-wave basis set with a 520 eV energy cutoff was used to represent valence electrons. Van der Waals interactions were included using Grimme's D3 correction.[39] To simulate the experimental electrochemical environment, electric fields ranging from -1.0 to +1.0 V/Å were applied to account for the effects of both potential and pH effects. The structures of the reaction intermediates were optimized for each electric field until the forces on the atoms were minimized to below 0.02 eV/Å. The most stable configurations were then chosen and used to calculate adsorbate energies at each applied field. The information on elementary steps and further computational and modelling details are provided in the **Supporting Information (SI)**.

*Materials* MWCNT (~20 nm diameter) obtained from CNano are purified before catalyst synthesis. Manganese(II) phthalocyanine (MnPc, >95%), iron(II) phthalocyanine (FePc, >95%), cobalt (II) phthalocyanine (CoPc, >95%), nickel(II) phthalocyanine (NiPc, >95%), and copper(II) phthalocyanine (CuPc, >95%) are obtained from PorphyChem Inc. and purified by triple-sublimation before catalyst synthesis. Hydrochloric acid (HCl, 37%), sulfuric acid ($H_2SO_4$, 98%), hydrogen peroxide ($H_2O_2$, 30%), potassium sulfate ($K_2SO_4$, >99%), potassium nitrate ($KNO_3$, >99%), potassium hydroxide (KOH, semiconducting grade, 99.99%), sodium hydroxide (NaOH, semiconducting grade, 99.99%), sodium nitroferricyanide (≥99%), sodium salicylate (≥99.5%), sodium hypochlorite solution (NaClO, reagent grade, 10-15% available chlorine), sulfanilamide (GR for analysis), N-(1-naphthyl)ethylenediamine dihydrochloride (>98%), *N, N'*-dimethylformamide (DMF, 99%), isopropanol (IPA, HPLC grade), Nafion 117 suspension (~5% in a mixture of lower aliphatic alcohols and water) were obtained from Sigma-Aldrich, and used without treatment. Nafion 117 membrane was obtained from Fuel Cell Store, and pre-activated using $H_2SO_4$ and $H_2O_2$ pre-treatment. Carbon cloth substrate is obtained from Sinerosz Technology (W1S1011). Deionized water (DI $H_2O$) is supplied by a MilliQ water system. Argon (Ar) and helium (He) gases (5.0 grade), and 5% hydrogen in argon mixture gas ($H_2$/Ar) are obtained from BOC Australia.

*Carbon nanotube purification* As-received MWCNT is purified by 30 min baking in the air at 400 °C, followed by bath sonication in 3 M HCl for 30 mins and further stirring in the acid for 6 hr at 80 °C. Solids are recovered by vacuum filtration and washed with DI $H_2O$ until the filtrate is near neutral. The MWCNT is further thermally treated at 1200°C in a 100 sccm 5% $H_2$/Ar flow for 2 hours.

*Catalyst preparation* MPc/CNT heterogeneous molecular catalysts are prepared by dissolving MPc (~2.5 mg) in 50 mL DMF, followed by adding 50 mg purified CNT. The mixture is bath-sonicated for 30 minutes before being further stirred at ambient conditions for 24 hours. The MPc/CNT catalysts are then recovered by filtration, washed with DMF and ethanol, and dried under vacuum.

*Material characterization* Powder X-ray diffraction (XRD) patterns are collected on a Stoe Stadi P diffractometer with Cu-Kα source (λ= 1.5406 Å). Spherical aberration-corrected high-angle annular dark field scanning transmission electron microscope (HAADF-STEM) images are recorded on an FEI Themis-Z microscope. The metal loading is determined using inductively coupled plasma optical emission spectroscopy

(ICP-OES, Avio 500, Perkin Elmer). X-ray photoelectron spectra (XPS) are collected on a Thermo Fisher Scientific K-Alpha+ spectrometer with an Al-Kα (1486.3 eV) source. All spectra were collected at a pass energy of 20 eV and a spot of 400 μm. X-ray absorption spectra (XAS) are acquired at the XAS beamline at the Australian Synchrotron and further analyzed by the Demeter Software package using FEFF 9.0 code. Absorption spectra are collected on a Shimadzu UV3600 UV-vis-NIR spectrometer. Gas product is analyzed using gas chromatography (GC, Shimadzu, GC2040) with a barrier discharge ionization (BID) detector and He as carrier gas.

***Electrochemical test*** Catalyst ink is prepared by suspending the catalyst in a 1/9 (v/v) water/iso-propanol solution containing 0.05wt% Nafion 117 at 5 mg mL$^{-1}$. The ink is deposited on a 1×1 cm$^2$ carbon cloth at 0.2 mg cm$^{-2}$ as the working electrode. A control electrode is further prepared by depositing purified CNT substrate on the carbon cloth at the same loading. Electrochemical tests are performed under a three-electrode configuration in an H-shaped two-chamber electrolyzer at 25 °C using an Autolab PGSTAT302N potentiostat. The two chambers are separated by a pre-activated Nafion 117 membrane. A pre-calibrated saturated calomel electrode (SCE) and a graphite rod are used as reference and counter electrodes, respectively. All potentials are reported against a reversible hydrogen electrode (V$_{RHE}$). A 0.1 M K$_2$SO$_4$ and 0.1 M KOH solutions containing 0.1 M KNO$_3$ are used as neutral and alkaline electrolytes, respectively. The electrolytes are pre-saturated by Ar bubbling for 30 minutes before NO$_3$RR is conducted.

***Product quantification***: Gas electrocatalysis products were analyzed by GC. The NH$_3$ quantification follows a reported salicylate method. Briefly, 500 μL of 0.4 M sodium salicylate in 0.32 M NaOH aqueous solution, 50 μL of a NaClO in 0.75 M NaOH solution (~5% active chlorine), and 50 μL of a sodium nitroferricyanide solution (10 mg mL$^{-1}$) are sequentially added to 3 ml of the electrolyte. After 1 hr of reaction, the optical absorbance at 675 nm is recorded. Ammonia in the cathode chamber and the tail gas scrubbing bottle are quantified. NO$_2^-$ is quantified by the Griess method. Briefly, 100 μL of a sulfanilamide aqueous solution (10 g L$^{-1}$ in 10 wt.% HCl) is added to 2 ml of the catholyte, followed by adding 100 μL of a 1 g L$^{-1}$ N-(1-naphthyl) ethylenediamine dihydrochloride solution, before the sample is quantified by optical absorbance at 540 nm. The Faradaic efficiency of H$_2$, NH$_3$ and NO$_2^-$ are calculated by the following equations:

$$FE_{H_2} = \frac{2xGPF}{iRT} \tag{1}$$

$$FE_{NO_2^-} = \frac{nxVF}{i} \tag{2}$$

$$FE_{NH_3} = \frac{nxVF}{i} \tag{3}$$

where $G$ is the Ar flow rate, $x$ is the H$_2$ fraction determined by GC or NO$_2^-$ and NH$_3$ concentration, $V$ is the catholyte chamber volume, $P$ = 101,325 Pa, $F$ is Faraday constant (96485 C mol$^{-1}$), $i$ is the current, $R$ is gas constant (8.314 J mol$^{-1}$), $n$ is the number of electrons for reducing NO$_3^-$ to NO$_2^-$ or NH$_3$.

Subsequently, the NH$_3$ partial current density ($j_{NH_3}$, mA cm$^{-2}$), and metal site specific turnover frequency for NH$_3$ ($TOF_{NH_3}$, s$^{-1}$) are calculated by:

$$j_{NH_3} = jFE_{NH_3} \tag{4}$$

$$TOF_{NH_3} = \frac{j_{NH_3}}{0.2 \times 10^{-3} n(M\%)/Mw_M} \tag{5}$$

where $j$ is the total current density, $m_{cat}$ is the mass of catalyst on the electrode, 0.2 is the mass loading of the catalyst (mg cm$^{-2}$), $M\%$ is the metal weight loading in the catalyst, and $Mw_M$ is the molar mass of the metal center.

**Acknowledgements**

This work was financially supported by JSPS KAKENHI (Grant Nos. JP23K13703 and JP24K17650), the Hirose Foundation, and the National Natural Science Foundation of China (Grant No. 22373097). Qiuling Jiang acknowledges the support from the China Scholarship Council. The authors are grateful to the Institute for Materials Research (IMR) at Tohoku University for providing computational resources through MASAMUNE-IMR (Project Nos. 202312-SCKXX-0203 and 202312-SCKXX-0207) and to the Institute for Solid State Physics (ISSP) at the University of Tokyo for the computational resources provided.

**Notes**

The authors declare no competing financial interest.